\documentclass{article}
\usepackage{spconf,amsmath,graphicx}
\usepackage[ruled, norelsize]{algorithm2e}
\usepackage{algorithmic}
\usepackage{amsmath} 
\usepackage{amssymb}

\usepackage{subcaption}
\usepackage{float}
\usepackage{color}

\makeatletter
\newcommand{\removelatexerror}{\let\@latex@error\@gobble}

\makeatother

\title{U-SLADS: UnSupervised Learning Approach for Dynamic Dendrite Sampling}
\name{Yan Zhang$^1$, Xiang Huang$^2$, Nicola Ferrier$^2$, Emine B. Gulsoy$^3$, Charudatta Phatak$^1$
}
\address{
$^1$Materials Science Division, Argonne National Laboratory, Lemont, IL \\
$^2$Mathematics and Computer Science Division, Argonne National Laboratory, Lemont, IL  \\
$^3$Department of Materials Science and Engineering, Northwestern Univeristy, Evanston, IL
}
\begin{document}
%\ninept
%
\maketitle
\begin{abstract}
Novel data acquisition schemes have been an emerging need for scanning microscopy based imaging techniques to reduce the time in data acquisition and to minimize probing radiation in sample exposure. Varies sparse sampling schemes have been studied and are ideally suited for such applications where the images can be reconstructed from a sparse set of measurements. Dynamic sparse sampling methods, particularly supervised learning based iterative sampling algorithms, have shown promising results for sampling pixel locations on the edges or boundaries during imaging. However, dynamic sampling for imaging skeleton-like objects such as metal dendrites remains difficult. Here, we address a new unsupervised learning approach using Hierarchical Gaussian Mixture Models (HGMM) to dynamically sample metal dendrites. This technique is very useful if the users are interested in fast imaging the primary and secondary arms of metal dendrites in solidification process in materials science.
\end{abstract}

\section{Introduction}
In most commonly used conventional point-wise imaging modalities, each pixel measurement can take up to a few seconds to acquire, which can translate to hours or even days for middle to high resolution image (e.g. 1024 $\times$ 1024 or 2048 $\times$ 2048 pixels) measurements. The sample exposure to a highly focused electron or X-ray beam for extended periods of time may furthermore damage the underlying object. Thus, minimizing the image acquisition time and radiation damage is of critical importance. Static sampling methods, such as random sampling, uniform spaced sampling and low-discrepancy sampling methods have been widely studied and used \cite{bib1}\cite{bib2}\cite{bib3}. Recently, sampling techniques where previous measurements are used to adaptively select new sampling locations have been presented. These methods, known as dynamic sampling methods, have been shown to significantly outperform traditional static sampling methods \cite{ds1}\cite{ds2}\cite{dilshan}\cite{yz}\cite{slads-net}.

SLADS \cite{dilshan} and SLADS-Net \cite{slads-net} are two dynamic sampling methods based on supervised learning approaches. The goal of SLADS and SLADS-Net is to select the measurements that minimizes the reconstruction error during the sampling process. In order to train SLADS/SLADS-Net, one needs corresponding pairs of representative extracted features and pre-calculated reconstruction errors from historical images. The mapping from features to reconstruction errors could be learned using various learning algorithms, such as linear regression, support vector regression or (deep) neural networks. In testing or experiment, SLADS/SLADS-Net will compute the reconstruction error as a score using extracted features for unmeasured pixel locations and select the one that has lowest score value for the next measurement. The key local descriptors for SLADS/SLADS-Net feature extraction are the gradients and variances close to edges in images. \cite{dilshan} and \cite{slads-net} showed that these algorithms successfully sampled the pixels on the "informative" boundaries of an object. The PSNR were high and the distortions were low between reconstruction and original images.

\begin{figure}[htb]
  \centering
  \centerline{\includegraphics[width=4cm]{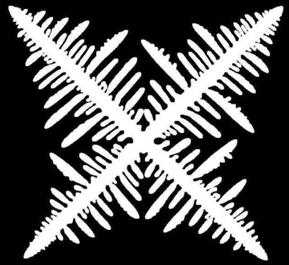}}
  %\centerline{(a) Result 1}\medskip
\caption{Metal dendrite image \cite{dendrite}. The measured pixel locations are used for Gaussian mixture model and the measured intensities are used for computing threshold. 
%\fix{XH: This seems to be a binary image. Do we only consider two-phase binary image? We may need to make it clear. But if it is binary, then I am not sure if Gaussian Mixture is a good model.}
}
\label{fig:dendrite}
\end{figure}

However, in the imaging fields, some researchers are more enthusiastic about the skeleton of an object instead of the edges. One important area is the metal solidification research, where researchers care more about the formation of metal dendrites during solidification \cite{dendrite}, as shown in Figure \ref{fig:dendrite}. An algorithm that can iteratively sample along the main direction of the object formation is of critical importance. Since it is relatively difficult to define the key features related to this phenomenon nor the metric for reconstruction, unsupervised learning approach is used to estimate the next measurement locations based on the calculated distribution using current measurements. Here, we use Hierarchical Gaussian Mixture Models (HGMM) for dynamic sampling and we name it U-SLADS because of its unsupervised learning fashion.

\section{U-SLADS Framework}
The core idea of U-SLADS is to use two-dimensional Gaussians to model the primary and secondary arms of dendrites. 
We discretize the 2D sample and then vectorize it as $X \in \mathbb{R}^N$, and define by $X_s$ the pixel value at location $s$. Assuming $t$ pixels have been measured, we can construct the current measurement vector as a combination of locations and intensities: %\fix{(strictly speaking, this is not a vector, but a $k \times 2$ matrix)},
$$
Y^{(t)} = 
\left[
\begin{array}{c}
s^{(1)}, X_{s^{(1)}} \\
\vdots \\
s^{(t)}, X_{s^{(t)}}
\end{array}
\right].
$$
Using the measurement vector $Y^{(t)}$, we then compute a threshold from the $t$ pixel intensities, and perform clustering of locations only for pixels with intensity larger than the threshold.

In U-SLADS, the key step is to select the next measurement pixel to update the measurement vector as
$$
Y^{(t+1)} = 
\left[
\begin{array}{c}
Y^{(t)} \\
%\vdots \\
s^{(t+1)}, X_{s^{(t+1)}}
\end{array}
\right]
$$
This process is repeated until the stopping criterion is met.

In SLADS and SLADS-Net, the dynamic sampling methods are based on supervised learning strategies which require training images and feature extractions. In each iteration of SLADS or SLADS-Net, local descriptors, such as gradients, variance and density, are computed to form feature vectors. Unmeasured locations having high feature scores are usually close to edges of training images and experimental objects. However, in dynamic sampling process for dendrites imaging, those features have little contribution to the measurement of dendrites formation of an object. Our proposed U-SLADS algorithm is thus based on unsupervised learning approach which iteratively update the distributions of measured locations and estimate the next measurement locations which might contribute to the existing clusters data distributions.

In U-SLADS, we use HGMM to select the next measurement locations in each iteration. We start with $5 \%$ random sampling and use Otsu' method \cite{otsu} to calculate a threshold for all measured intensities. We then apply GMM \cite{bishop} clustering on all measurement locations which have intensities larger than the threshold. For the unmeasured locations in each cluster, we calculate the weighted distances (Mahalanobis distances) to their cluster centroids. We later perform measurements on the unmeasured locations of top n distances (n is user's choice). We apply the same procedure in a hierarchical fashion for each cluster until only one single cluster is found.

\begin{figure}[htb]
  \centering
  \centerline{\includegraphics[width=1.1\linewidth]{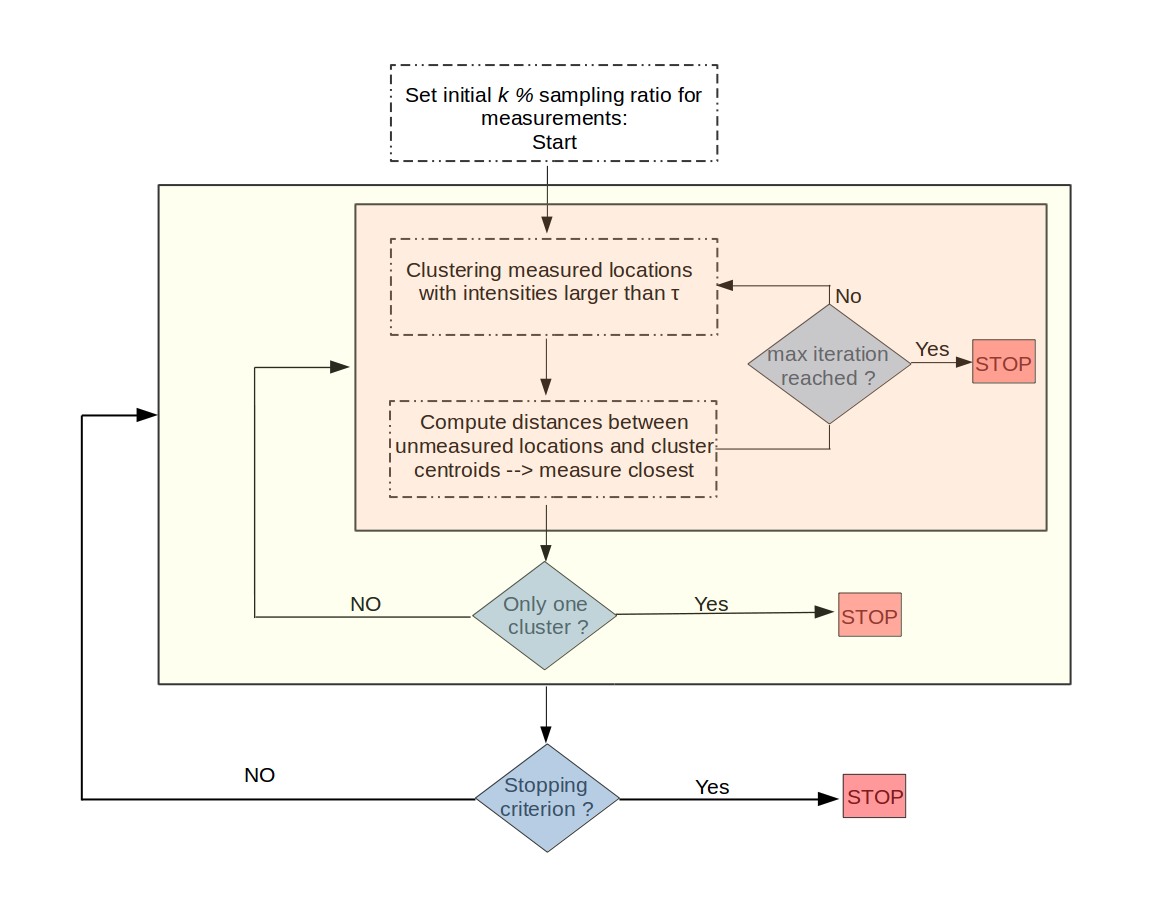}}
  %\centerline{(a) Result 1}\medskip
\caption{U-SLADS Framework. The stopping criterion of the orange box is controlled by maximum iteration in each GMM layer. The iteration of the yellow box stops when only one single cluster is found. The outer loop is controlled by the sampling ratio for the total measurements.}
\label{fig:framework}
\end{figure}

\section{U-SLADS Algorithms}
We describe U-SLADS as three parts: the main function, layer-wise HGMM function and Bayesian Information Criterion (GMMbic) \cite{bic} function.

\subsection{U-SLADS main function}
The main function of U-SLADS is to store the layer-wise measurements of the clusters in each layer. In each iteration, it calls $layerGMM$ function which is used to calculate the locations to be measured and perform the measurement. Then the updated number of clusters of the current updated measurement vector $Y^{(t+1)}$ is evaluated. If there's only one cluster, then $Y^{(t+1)}$ is pushed into a Stack which is a record of the measurements in each hierarchical layer of GMM models. If there exists multiple clusters, then $Y^{(t+1)}_k$ for each cluster $k$ is pushed into a Queue which will be used for next layer GMM computation. If the Queue is not empty, then a sub-image is constructed using the measurement vector $Y^{(t+1)}_k$ pop out from the Queue and repeat the whole process of the main function.  

{\small
\begin{figure}[h]
 \removelatexerror
\begin{algorithm}[H]
   \caption{U-SLADS main}
 %\KwData{this text}
 %\KwResult{how to write algorithm with \LaTeX2e }
 {\bf Function} $Y^{(t+1)} \leftarrow USLADS(X^{(t)})$ \\
 Queue = [ ], stack = [ ]\;
 \While {($sampling\_ratio <= \phi\%$)}{
  $X^{(t+1)} = layerGMM(X^{(t)})$\;
  $Y^{(t+1)} = Y^{(t)} \cup \left[ {s^{(t+1)}, X_{s^{(t+1)}}} \right]$\;
  \eIf{G.components == 1}{
   stack.push($Y^{(t+1)}$)\;
   %current section becomes this one\;
   }{
   
   \For{k=1:G.components}{
   Queue.push($Y^{(t+1)}_{k}$)\; \hspace{1em} %// $k \in G.components$
   }
  }
  \eIf{Queue not empty}{
  $Y^{(t+1)}_{i}$ = Queue.pop()\;
  $X^{(t+1)}_{i} = ConstructImage(Y^{(t+1)}_{i})$\;
  }{
  break\;
  }
 }
return $Y^{(t+1)}$
 %\caption{USLADS Algorithm}
 
\label{algo:main}
\end{algorithm}
\end{figure}
}

\subsection{Layer-wise HGMM function}
The layer-wise HGMM function first computes a threshold $\tau$ using all measured intensities by Otsu's method \cite{otsu}. The measurement locations in the measurement vector $Y^{(t)}.s$ having intensity values larger than the threshold are preserved and used by GMMbic function to calculate the mixture of Gaussian distributions \cite{bishop}. The unmeasured locations $u^{(t)}$ are then used to calculate their predicted cluster labels using the GMM of current step. The unmeasured locations in each assigned cluster $\hat{u}_{k}^{(t)}$ are used to compute their weighted distance from cluster centroids using the mean vectors $\mu_k$ and covariance matrices $\Sigma_k$. The weighted distances $D$ are sorted and the unmeasured locations having the first $\epsilon$ closest distances will be measured.

{\small
\begin{figure}[h]
 \removelatexerror

\begin{algorithm}[H]
   \caption{Layer-wise HGMM}
 %\KwData{this text}
 %\KwResult{how to write algorithm with \LaTeX2e }
  {\bf Function} $X^{(t+1)} \leftarrow layerGMM(X^{(t)})$ \\
 
   %\eIf{$max(X^{(t)}) < 1$}{
   %$\tau = 0$\; 
   %}{
   %$\tau = otsu(X^{(t)})$\;
   %}

 \For{i = 1 : maxiter}{
 	$\tau = otsu(X^{(t)})$\;
 	$Y^{(t)} = Segment(X^{(t)} >= \tau)$ \;
    
    %$Y^{(t)} = UpdateIndex(X^{(t)})$\;
    $G^{(t+1)} = GMMbic(Y^{(t)}.s)$\;
    $labels = G^{(t+1)}.predict(u^{(t)})$\;
    $\hat{u}_{k}^{(t)} = label\_sort \left( u^{(t)} \right)$\;
    \For {k = 1 : K}{
    	$D = \sqrt{ \left( \hat{u}_{k}^{(t)} - {\mu_k} \right)^T {\Sigma_k}^{-1} \left( \hat{u}_{k}^{(t)} - {\mu_k} \right) }$\;
    }
    
    $idx = argsort(D)$\;
    
    \eIf{$len(D) < \epsilon$}{
    	$X^{(t+1)} = PerformMeasurement(\hat{u}^{(t)}, idx)$\;
    }{
    	$X^{(t+1)} = PerformMeasurement(\hat{u}^{(t)}, idx[1:\epsilon])$\;
    }

    %\If{$G^{(t)}.\mu - G^{(t-1)}.\mu < \epsilon$}{
    %	break
    %}
    
    %$X^{(t+1)} = PerformMeasurement(G^{(t+1)}.\mu)$\;   
 }
return $X^{(t+1)}$ 

\label{algo:gmm}
\end{algorithm}
\end{figure}
}

\subsection{GMMbic function}
The GMMbic function is to calculate the BIC scores \cite{bic} for different hypothesized number of clusters. The number of clusters having the lowest BIC score will be the GMM model parameter to be used in the current step. A binary search method may be used to improve the search efficiency.

\section{Hyper-parameters}
A total of four hyper-parameters can be tuned. The \\$sampling\_ratio$ in Algorithm \ref{algo:main} is the stopping criterion regards to the percentage of dynamic sampling. The $maxiter$ in Algorithm \ref{algo:gmm} indicates the number of iterations in each layer of GMM and $\epsilon$ defines the maximum allowance of the locations to be measured in each GMM run. The $n$ in Algorithm \ref{algo:bic} is set to be maximum number of clusters in BIC score search.

\section{Results and Conclusion}
\begin{figure}[h]
%\centering
        \begin{subfigure}[htbp]{0.25\textwidth}
                \includegraphics[width=\linewidth]{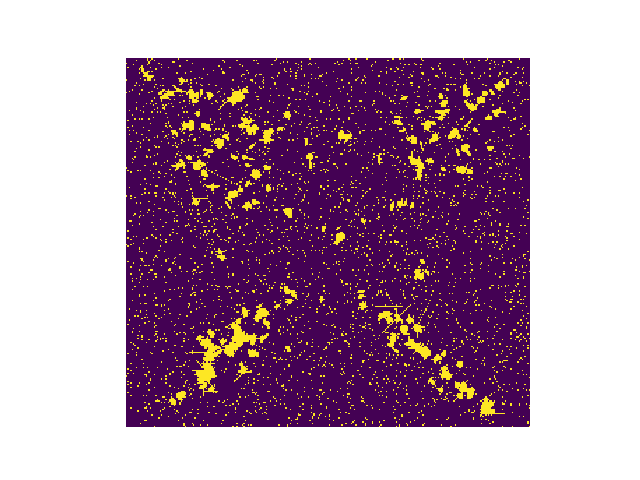}
                \vspace{-0.9cm}
                \caption{10\% sampled mask.}
                \label{fig:2d}
        \end{subfigure}%
        \hspace{-3em}
		\hfill
                \begin{subfigure}[htbp]{0.25\textwidth}
                \includegraphics[width=\linewidth]{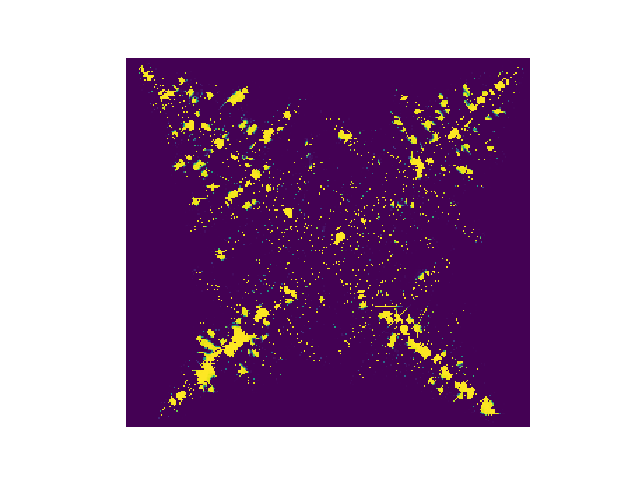}
                \vspace{-0.9cm}
                \caption{10\% sampled image.}
                \label{fig:2d}
        \end{subfigure}%
        \hfill 
        \begin{subfigure}[htbp]{0.25\textwidth}
                \includegraphics[width=\linewidth]{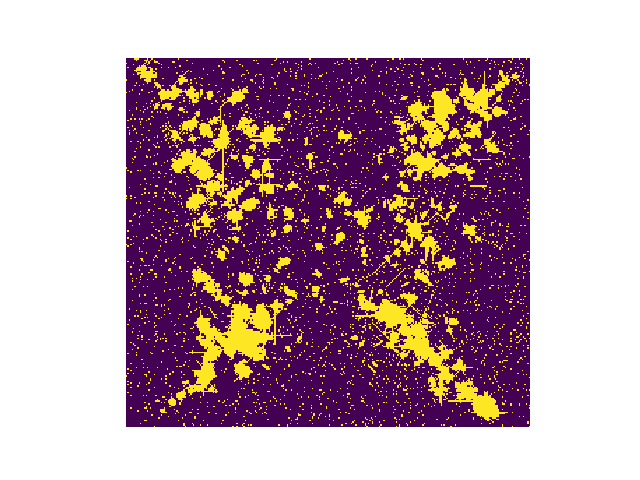}
                \vspace{-0.9cm}
                \caption{20\% sampled mask.}
                \label{fig:10d}
        \end{subfigure}  
        \hspace{-3em}
        \hfill
         \begin{subfigure}[htbp]{0.25\textwidth}
                \includegraphics[width=\linewidth]{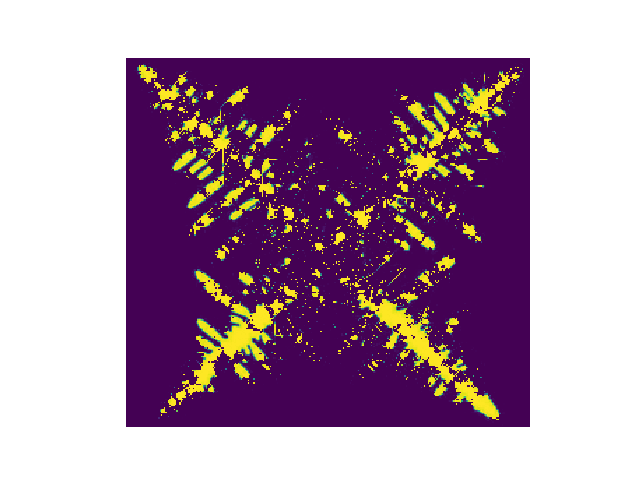}
                \vspace{-0.9cm}
                \caption{20\% sampled image.}
                \label{fig:10d}
        \end{subfigure} 
      
        \hspace{-3em}
        \hfill
        
		\begin{subfigure}[htbp]{0.25\textwidth}
                \includegraphics[width=\linewidth]{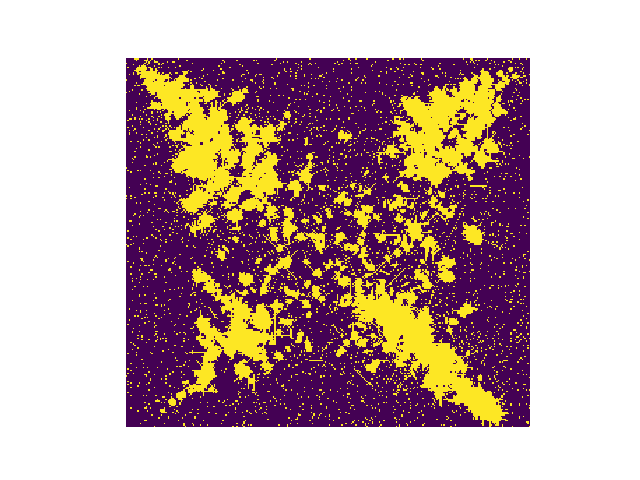}
                \vspace{-0.9cm}
                \caption{30\% sampled mask.}
                \label{fig:10d}
        \end{subfigure} 
        \hspace{-3em}
        \hfill
        \begin{subfigure}[htbp]{0.25\textwidth}
                \includegraphics[width=\linewidth]{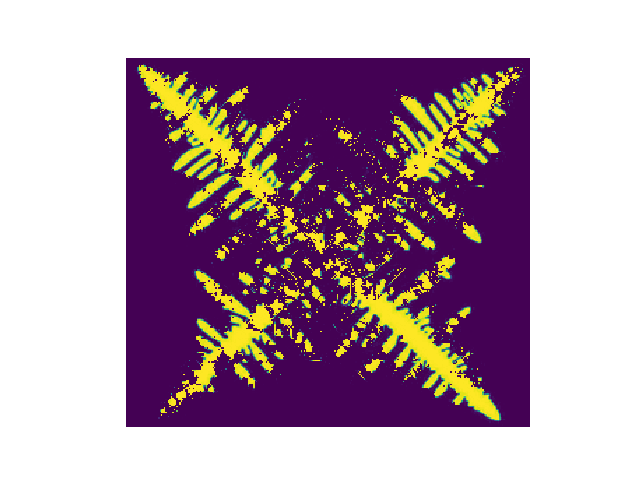}
                \vspace{-0.9cm}
                \caption{30\% sampled image.}
                \label{fig:10d}
        \end{subfigure} 
        \hspace{-3em}
         \hfill  
        \begin{subfigure}[htbp]{0.25\textwidth}
                \includegraphics[width=\linewidth]{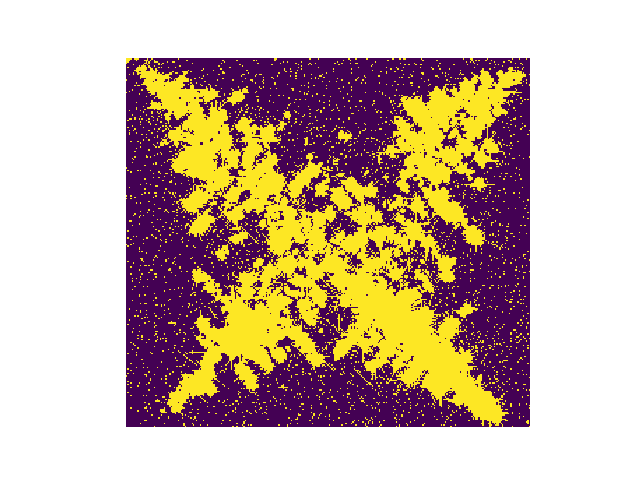}
                \vspace{-0.9cm}
                \caption{40\% sampled mask.}
                \label{fig:10d}
        \end{subfigure} 
        \hspace{-3em}
        \hfill
        \begin{subfigure}[htbp]{0.25\textwidth}
                \includegraphics[width=\linewidth]{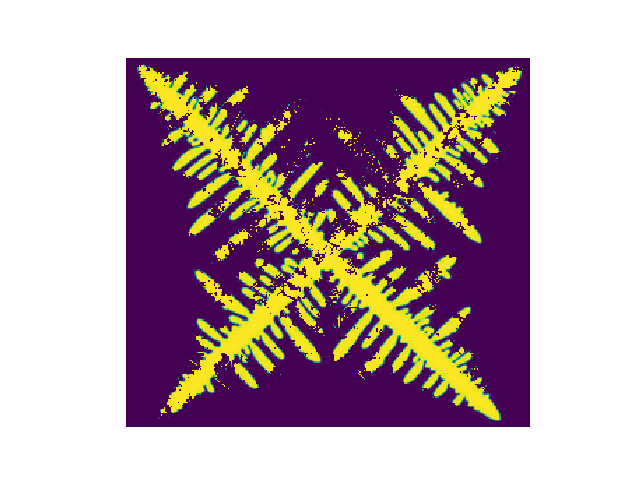}
                \vspace{-0.9cm}
                \caption{40\% sampled image.}
                \label{fig:10d}
        \end{subfigure}

        \caption{Sampled masks and images at different sampling ratios of U-SLADS. Left column is the sampled mask (measured locations) and right column is the corresponding sampled image (measured intensities).}
        \label{fig:result}
\end{figure}

We applied U-SLADS algorithm on a simulated dendrite image as shown in Figure \ref{fig:dendrite}. We used $maxiter=10$ and $\epsilon=10$, started with $5\%$ random sampling initially, and selected the future measurements iteratively using dynamic sampling up to $sampling\_ratio = 40\%$. Figure \ref{fig:result} shows the resulted sampled masks (measured locations) and sampled images (measured intensities). We can see that at $10\%$, the algorithm coarsely estimated the Gaussian distribution of the four primary arms and only a few secondary arms of the metal dendrites. Since then, U-SLADS started to find the sub-sets of Gaussian distributions that belongs to secondary arms in a hierarchical fashion. At $40\%$, U-SLADS has almost found all the feature distributions of the dendrites.

\begin{figure}[h]
  \centering
  \centerline{\includegraphics[width=7cm]{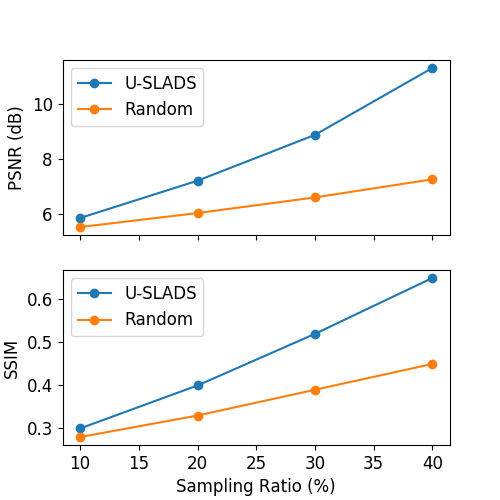}}
  %\centerline{(a) Result 1}\medskip
  \vspace{-0.2cm}
\caption{Peak Signal-to-Noise Ratio (PSNR) and Structural Similarity (SSIM) at different sampling ratios using sampled image without reconstruction.}
\label{fig:psnr}
\end{figure}

\begin{figure}[h]
  \centering
  \centerline{\includegraphics[width=7cm]{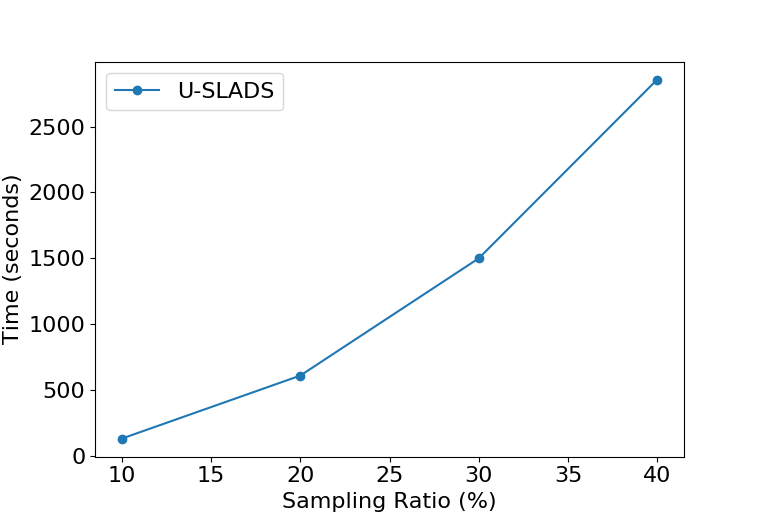}}
  %\centerline{(a) Result 1}\medskip
  \vspace{-0.2cm}
\caption{Time used at different sampling ratios.}
\label{fig:time}
\end{figure}

Our U-SLADS algorithm is a novel strategy for dynamic dendrite sampling. It outperforms the traditional random sampling method as shown in Figure \ref{fig:psnr} (U-SLADS has a PSNR of 11.31 dB over 7.25 dB by random sampling and structural similarity of 0.65 over 0.46 at 40\% measurements) and provides an alternate when boundary-focused dynamic sampling is not applicable. U-SLADS is an unsupervised learning based approach so training is not required. One limitation of U-SLADS is that the computation time increases exponentially with the sampling ratio, as shown in Figure \ref{fig:time}.

{\small
\begin{figure}[h]
 \removelatexerror

\begin{algorithm}[H]
   \caption{GMMbic}
 %\KwData{this text}
 %\KwResult{how to write algorithm with \LaTeX2e }
  {\bf Function} $G^{(t)} \leftarrow GMMbic(Y^{(t)})$ \\
 bic = [ ]\;
 \For{i = 1 : n}{
 	\If{$Y^{(t)}.size() > n$}{
    	$G^{(t)} = GaussianMixture(Y^{(t)}.s, i)$\;
        bic.push(G.score)\;
    }
 }
 \eIf{$len(bic) > 0$}{
 	k = argmin(bic) + 1\;
 }{
 	k = 1\;
 }
 $G^{(t)} = GaussianMixture(Y^{(t)}.s, k)$\;

return $G^{(t)}$

\label{algo:bic}
\end{algorithm}
\end{figure}
}

\section{Acknowledgments}
This material is based upon work supported by Laboratory Direct Research and Development (LDRD) funding from Argonne National Laboratory, provided by the Director, Office of Science, of the U.S. Department of Energy under Contract No. DE-AC02-06CH11357.

\end{document}